\documentclass[useAMS,usegraphicx,usenatbib]{mn2e}

\newcommand{\aap}{A\&A}
\newcommand{\mnras}{MNRAS}
\newcommand{\apj}{ApJ}
\newcommand{\apjl}{ApJ}
\newcommand{\app}{APh}
\newcommand{\nn}{\nonumber}
\def\lesssim{\buildrel < \over {_{\sim}}}
\def\gtrsim{\buildrel > \over {_{\sim}}}

\title[X-ray filaments in SN1006]{Spatial structure of X-ray filaments in SN
1006}
\author[G. Morlino, E. Amato, P. Blasi and D. Caprioli]{
G. Morlino$^{1,2,3}$\thanks{E-mail: morlino@arcetri.astro.it}, 
E. Amato$^{1,3}$  \thanks{E-mail: amato@arcetri.astro.it}, 
P. Blasi$^{1,3}$\thanks{E-mail: blasi@arcetri.astro.it; blasi@fnal.gov}, and 
D. Caprioli$^{1,2,3}$\thanks{E-mail: caprioli@arcetri.astro.it}, \\
$^{1}$INAF/Osservatorio Astrofisico di Arcetri, Firenze, Italy\\
$^{2}$Physics Department, North Carolina State University, Box
8202, Raleigh, NC 27695 \\
$^{3}$Kavli Institute for Theoretical Physics, Kohn Hall Santa
Barbara, CA 93106-0001, USA}

\begin{document}

\date{Accepted 2010 March 10.  Received 2010 March 10; in original form 2009
December 15}


\maketitle

\label{firstpage}

\begin{abstract}
The theory of Non-Linear Diffusive Shock Acceleration (NLDSA) predicts the
formation of a precursor upstream of the shock, where accelerated particles
diffuse and induce magnetic field amplification through streaming instability.
The non detection of this precursor in X-rays in {\it Chandra} observations of
the north-eastern region of SN 1006 (G329.6+14.6) led to impose an upper limit
to the X-ray emission generated by accelerated electrons diffusing in this
precursor, at an emissivity level of $<1.5$ per cent of the emission from the
downstream region \citep{Long03}. This has been used as an argument against
Fermi acceleration at this shock. Here we calculate the spectrum and spatial
distribution of accelerated particles in SN 1006 and show that {\it Chandra}
results (including more recent data) are in perfect agreement with the
predictions of NLDSA suggesting efficient particle acceleration and magnetic
field amplification upstream of the shock by a factor $\sim 10$.
\end{abstract}


\section{Introduction}\label{sec:intro}

The blast waves produced by supernova remnants (SNRs) expanding in the
interstellar medium (ISM) are thought to be the sites where the bulk of Galactic
cosmic rays (CR) are accelerated. The simplest conversion mechanism of kinetic
energy of the expanding plasma into accelerated particles is Diffusive Shock
Acceleration (DSA). Narrow rims of non-thermal X-ray emission, associated with
the outer shocks of several young SNRs \citep[see e.g.\ ][and references
therein]{par06}, are commonly interpreted as due to synchrotron emission
radiated by $\sim$ TeV electrons \citep[for the case of SN 1006 see
][]{bamba03}. These X-ray filaments might also represent an indirect evidence of
efficient acceleration of ions at the same shocks, since their thickness is most
easily explained in terms of severe synchrotron losses of electrons in intense
magnetic fields  \cite[in the range 100-500 $\mu$G,][]{ballet06}, to be compared
with the ISM magnetic field, $\sim 3 \mu$G. 
In order to account for this difference, it has been proposed that magnetic
fields are amplified in the shock region by the streaming of cosmic ray ions.
This gives rise to an interesting scenario in which magnetic fields are
generated by cosmic rays while in turn making CR acceleration more efficient in
a complex chain of non linear effects. If indeed the magnetic field
amplification (MFA) is due to CR streaming, this may happen either via resonant
\cite[]{skilling} or non-resonant instabilities \cite[]{bell04}, though many
variations on the theme are possible. The peculiarity of CR induced MFA is that
it occurs upstream of the shock thereby reducing the acceleration time. On the
other hand MFA could also be generated downstream by fluid instabilities 
\citep[e.g.\ ][]{giajo07} induced by the shock corrugation while propagating in
a inhomogeneous medium. Finally the narrow filaments might be due to damping of
the magnetic field \cite[]{Pohl05} rather than to severe synchrotron losses,
although correspondingly thin filaments in the radio emission do not seem to be
observed \cite[]{roth}. The only way to discriminate among these different
scenarios is through extensive comparison between theory and data.

The spatial profile of the X-ray brightness around the shock provides
invaluable information on the acceleration mechanism and the magnetic field
generation. Based on the non detection of X-rays with {\it Chandra} from the
upstream region of the north-eastern limb of SN 1006, \cite{Long03} concluded
that any X-ray halo upstream must be at least a factor $\sim 70$ fainter than
the peak brightness of the shell. Based on this finding, these authors suggested
that Fermi acceleration is not the acceleration mechanism at work or that the
jump in magnetic field at the shock is $\gg 4$. They also question the
possibility that observations may be explained by the non linear theory of DSA. 

In this Letter we apply the theory of NLDSA to show that its predictions
provide an excellent description of {\it Chandra} observations, which in turn
may be used to constrain the acceleration efficiency in SN 1006. 
A previous attempt at using the same technique was made in \cite[]{bere2003},
where a different filament was studied and some aspects of the theory were
treated in a more simplified way (for instance self-consistent, space dependent
MFA in the precursor was not accounted for).

\section{NLDSA model for SN 1006}\label{sec:theory}


In NLDSA theory, the overall shock structure and the outcome of the particle
acceleration process are inextricably linked. When acceleration is efficient,
the pressure of accelerated particles affects the shock dynamics, leading to the
formation of a precursor, namely a region where the fluid velocity 
progressively decreases while approaching the shock from far upstream. This
causes the total compression factor between upstream infinity and downstream to
exceed 4. At the same time the streaming of accelerated particles is responsible
for instabilities that lead to MFA. In turn, the fluid profile in the precursor
and the amplified, turbulent magnetic field, with the scattering it provides
(and possibly the induced energy losses), determine the efficiency of particle
acceleration and the resulting spectrum, including its high energy cutoff.

Here the shock structure and the accelerated particle spectrum are computed as in
\cite{mor09}: the  basic structure of the calculation is that proposed by
\cite{amato05,amato06}, but the conservation equations are modified so as to
also take into account the dynamical reaction of the self-generated 
magnetic field \citep[]{cap09}. The compressibility of the overall fluid is
deeply affected by the  magnetic contribution as soon as the magnetic field
pressure becomes comparable to the thermal pressure upstream. The net result is
that the total compression ratio between downstream and  upstream infinity,
$R_{tot}=u_0/u_2$, decreases with respect to the case in which this dynamical
backreaction is neglected, thereby leading to a smoother precursor and
accelerated particle spectra that are less concave and closer to power-laws.

The normalization of the proton spectrum is an output of our non linear
calculation, once a recipe for injection in the acceleration process is
established. Here we follow \cite{bgv05}, and assume that the fraction of
injected particles, $\eta_{\rm inj}$, is
\begin{equation}
\eta_{\rm inj}= 4/\left(3\sqrt{\pi}\right) 
(R_{\rm sub}-1) \, \xi^3\,e^{-\xi^2}\,.
\label{eq:inj}
\end{equation} 
where $R_{sub}=u_1/u_2$ is the compression ratio at the subshock and $\xi\sim
2-4$ is the ratio between the injection momentum and the momentum of thermal
particles downstream, $p_{\rm th,2}$. While $\xi$ parametrizes the poorly known
microphysics of the injection process, $p_{\rm th,2}$ is an output of the
calculation: as a result, the injection efficiency is affected by the dynamical
reaction of both particles and fields.

Energy losses are not important for accelerated protons, so that their maximum
energy is obtained by equating the acceleration time and the age of the
remnant, $t_{acc}(p_{p,\rm max})=t_{\rm SNR}$ \cite[]{dammax}. For electrons
the maximum energy is reached when the acceleration time equals the minimum
between the age of the remnant and the loss time scale, $\tau_l$, weighed by the
residence times, $t_r$, upstream and downstream of the shock (subscripts 1 and 2
respectively), over one cycle:
\begin{equation} \label{eq:pemax}
 t_{acc}(p) =
 \left[t_{r,1}(p)+t_{r,2}(p)\right]\left[{\frac{t_{r,1}(p)}{\tau_{l,1}(B_1,p)}}+
 {\frac{t_{r,2}(p)}{\tau_{l,2}(B_2,p)}}\right]^{-1}\ .
\end{equation}
Once $t_{r,1}$ and $t_{r,2}$ are written explicitly in the context of NLDSA
\citep[from Eqs.~(25) and (26) of][]{dammax}, Eq.~(\ref{eq:pemax}) must be
solved numerically for $p_{e,\max}$. The following approximate analytical
solution can be found when only synchrotron losses are relevant 
\citep[Eq.~(4)]{mor09}:
\begin{equation} \label{eq:Emax}
 p_{e,\max} = 137 \,H(p_{e,\max})\,
   \left(\frac{B_1}{\mu \rm G}\right) ^{-\frac{1}{2}}
   \left(\frac{u_0}{3000\,\rm km/s}\right) \rm \frac{TeV}{c}\,.
\end{equation}

Here $H(p)$ is a function of the shock modification taking into account the
mean plasma speed, $u_p=u_0 U_p(p)$, that a particle with momentum $p$
experiences in the precursor (see Eq.~(8) of \cite{amato05} and Eq.~(4) of
\cite{mor09}). Here $H(p)$ is normalized in such a way that $H(p)=1$ in the
test-particle limit.

The electron spectrum at the shock, $f_{e,0}(p)$, follows the proton spectrum
for $p\ll p_{e,\rm max}$, provided the diffusion coefficient is the same for
both species. The normalization of the electron spectrum relative to the proton
spectrum, $K_{ep}$, is instead unconstrained from the dynamics, since electrons
have no dynamical role, and can only be derived from observations.

The electron spectrum at energies around and above $p_{e,\rm max}$, namely the
shape of the cutoff, has never been computed in the context of NLDSA. Since the
spectra we find for electrons at $p<p_{e,\rm max}$ are not far from being power
laws with slope $\sim 4$, we adopt the modification factor calculated by
\cite{zir07} for strong shocks in the test particle regime. The resulting
electron spectrum at the shock, in the loss dominated case, is:
\begin{equation} \label{eq:f_e(p)}
f_{e,0}(p)= K_{ep}\,f_{p,0}(p)
 {\left[1+0.523 \left(p/p_{e,\rm max}\right)^{\frac{9}{4}}\right]}^2
 e^{-p^2/p_{e,\rm max}^2} \,.
\end{equation} 
An important point to notice in Eq.~(\ref{eq:f_e(p)}) is that the cutoff is
super-exponential, which reflects in the shape of the synchrotron spectrum,
making it different from what is assumed in most of the previous work on the
subject. On the other hand, if the maximum momentum is determined by the age of
the remnant, then the cutoff shape is a simple exponential. 



While a few different mechanisms have been proposed for MFA at SNR shocks,
here  we focus on the waves that are resonantly excited \cite[]{skilling}
upstream  by the streaming of cosmic rays accelerated at the shock. While
streaming cosmic rays can excite very effectively also non-resonant modes
\citep[]{bell04}, these are likely dominant only at early times, while after
the beginning of the Sedov phase MFA is mostly due to resonant waves
\citep[]{kinetic}. Moreover the role of these non resonant modes for
scattering the particles is yet to be  clarified.

When the predictions of linear theory are extrapolated to the non-linear
regime of MFA (which one is forced to do for lack of a better treatment), the
resulting field strengths are in agreement with the values inferred by
identifying the thickness of the X-ray filaments with the synchrotron loss
length of the highest energy electrons. The strength of the magnetic field at
the position $x$ upstream, $\delta B(x)$, in the absence of damping, can be
estimated from the saturation condition. For modified shocks this reads
\citep[Eq.~(43)]{cap09}:
\begin{equation} \label{eq:B_amp}
\frac{\delta B(x)^2}{8\pi\rho_0u_0^2} = U(x)^{-3/2}
\left[\frac{1-U(x)^2}{4\,M_{A,0}} 
           \right] \,,
\end{equation}
where the {\it lhs} is the turbulent magnetic pressure normalized to the
incoming momentum flux at upstream infinity, $U(x)=u(x)/u_0$ is the normalized
velocity profile in the precursor, and $M_{A,0}=u_0/v_A$ with $v_A$ the
Alfv\'en velocity at upstream infinity, where only the background magnetic
field, $B_0$, assumed parallel to the shock normal, enters.
Eq.~(\ref{eq:B_amp}) correctly describes the effect of compression in the
shock precursor through the term $U(x)^{-3/2}$.

The magnetic field downstream of the subshock is further enhanced by
compression, so that  $B_2= R_{B} \, B_1$, where $B_1$ is the magnetic field
immediately upstream of the subshock and, for Alfv\`en waves (turbulence
perpendicular to the shock normal), $R_B=R_{sub}$. We neglect  any kind of
magnetic damping both in the upstream and in the downstream region.

\section{The size of the X-ray filaments} \label{sec:thickness}

SN 1006 is one of the best suited remnants for the study of the spatial profile
of X-ray emission since its non-thermal rims, with a typical thickness of
30''-40'', are well resolved by {\it Chandra}, whose angular resolution is
$\Delta\phi \sim 1''$. The spatial resolution corresponding to this value of
$\Delta\phi$, $\Delta x= 3 \times 10^{16} \left(d/2\,{\rm kpc} \right)
\left(\Delta\phi/{\rm arcsec} \right) {\rm cm}$ on a source at a distance $d$,
has to be compared with the extension of the region in the upstream (downstream)
fluid where electrons radiate at the observed frequency $\nu_0$, i.e. $\Delta
R_{1(2)}(\nu_0)$. Before illustrating the results of the detailed calculation,
we provide some simple but useful approximate relations. 

In the upstream region electrons diffuse up to a distance $L_{\rm diff}=D(p)/u_p(p)$ from the shock, where $D$ is the diffusion coefficient and $u_p$ is the effective fluid velocity experienced by particles with momentum $p$. The spatial distribution of electrons, $f_1(p)$, is roughly constant up to the diffusion length $L_{\rm diff}$, and is then cut off \cite[]{Blasi10}. Assuming Bohm diffusion, $D_B=pc^2/(3eB)$, we find:
\begin{eqnarray} \label{eq:Ldiff1}
 \Delta R_1(\nu)  \simeq L_{\rm diff}(h \nu) = \frac{D_B(p)}{u_{p}(p)} 
    = \frac{p c^2}{3 e B_1 u_p(p)} = \hspace{1.6cm}\nn \\
    9 \times 10^{17} \, 
    \left( \frac{h \nu}{1 \, \rm keV} \right)^{\frac{1}{2}} 
    \left( \frac{B_1}{10 \,\rm \mu G} \right)^{-\frac{3}{2}}
    \left( \frac{u_0\,U_p(p)}{3000 \,\rm km/s} \right)^{-1} \, {\rm cm},
\end{eqnarray} 
where we used the fact that synchrotron emission of electrons with energy $E$ in
a $\mu G$ field peaks at energy $h \nu= 1.5 \times 10^{-2} B_{\rm \mu G}\,
{E_{\rm TeV}}^2 {\rm eV}$. Since $L_{\rm diff}$ strongly depends on the magnetic
field strength, the possibility of resolving the upstream X-ray halo depends on
the level of upstream MFA.

In the upstream region, electrons with $E=E_{\max}$ radiate at a typical frequency
\begin{equation} \label{eq:nu_1}
 h\nu_1 = 0.28 \,H(p_{e,\max})(u_0/3000 {\rm km\ s^{-1}})^2\ \rm keV\,, 
\end{equation} 
where we used the expression in Eq.~(\ref{eq:Emax}) for $E_{\max}$.
Clearly Eq.~(\ref{eq:Ldiff1}) holds only for $E<E_{max}$, since at
higher energies the diffusion region upstream becomes regulated by synchrotron
losses:
\begin{equation}
\Delta R_1(\nu>\nu_1)\simeq L_{\rm diff}(h \nu_1)\ .
\label{eq:dr1b}
\end{equation}
The size of the radiating region downstream is determined by the competition
between advection and diffusion of the electrons during their life-time, namely
$L_{\rm adv}^{\rm loss}= u_2\, \tau_{l}(p)\propto p^{-1}$ and $L_{\rm diff}^{\rm
loss}=\sqrt{D(p) \, \tau_{l}(p)}$ (the latter is
independent of energy for Bohm diffusion). These two length-scales are equal for
particle momenta corresponding to photon energy:
\begin{equation} \label{eq:nu_rim}
 h \nu_2= 1.00 \left( \frac{u_0}{3000\, \rm km\ s^{-1}} \right)^2 
  \left( \frac{R_{tot}}{4} \right)^{-2}  {\rm keV}\ .
\end{equation}
At frequencies lower than $\nu_2$:
\begin{equation} \label{eq:dr2a}
 \Delta R_2(\nu<\nu_2)\simeq L_{\rm adv}^{\rm
  loss}=\frac{u_0}{R_{tot}}\tau_l(p)\propto \nu^{-1/2}\ ,
\end{equation}
while for $\nu>\nu_2$ the downstream emission comes from a region of constant size:
\begin{equation} \label{eq:dr2b}
 \Delta R_2(\nu>\nu_2)\simeq L_{\rm diff}^{\rm loss}
  = 4.5 \times 10^{17} \left( \frac{B_2}{40 \rm \mu G} \right)^{-\frac{3}{2}}
  {\rm cm}\ .
\end{equation}

In order to compare $\Delta R_1$ and $\Delta R_2$ with the observed rim
thickness, we need to take into account projection effects. Let us define the
observed thickness, $\Delta R_{1(2)}^{\rm obs}$, as the width over which the
emissivity drops to half the peak value. Assuming spherical geometry, an
emission profile $I_\nu\propto \exp[-x/\Delta R]$ and $\Delta R_{1(2)}$ much
smaller than the SNR radius, the observed widths are $\Delta R_1^{\rm obs}= 0.7
\Delta R_1$ and $\Delta R_2^{\rm obs}= 4.6 \Delta R_2$. Hence projection
effects make the observed width of the downstream emission region appear larger
than the physical one.

When we observe the filaments at frequencies $\nu> \nu_1, \nu_2$, the ratio
between the observed thickness downstream and upsteam can be estimated as:
\begin{equation}
 \frac{\Delta R_{2}^{\rm obs}}{\Delta R_{1}^{\rm obs}} =
  \frac{4.6 \, L_{\rm diff}^{\rm loss}}{0.7 \, L_{\rm diff}(p_{e,\rm max})}
  = 12 \, \sqrt{\frac{1+R_B R_{tot} U_p}{R_B^3(1-R_{tot}^{-1}U_p^{-1})} }\ ,
\end{equation}
where the last equality is obtained using the full expression for $H(p)$ that
appears in Eq.~\ref{eq:pemax} \citep[see Eq.~(4) of][]{mor09}. Remarkably this
result is independent of magnetic field strength, shock speed and electron
energy, while it only depends on the compression ratios (magnetic compression at
the subshock $R_B$ and total compression $R_{tot}$) and on $U_p=u_{p}/u_{0}$.
Since $R_{sub}/R_{tot}\leq U_p\leq 1$ and $\sqrt{11} \leq R_B\leq 4$, using
reasonable values for $R_{tot}$ the ratio  $\Delta R_2^{\rm obs}/\Delta R_1^{\rm
obs}$ is always between 6 and 8. Since the rim in SN 1006 has an overall width
of 30''-40'' for $\nu \gtrsim \nu_1,\nu_2$, the result above implies that
the upstream halo should have an observed width of $\lesssim 6''$. 

\section{Comparison between theory and observations}\label{sec:obs}

\begin{figure}
\begin{center}
{\includegraphics[angle=0,width=.48\textwidth]{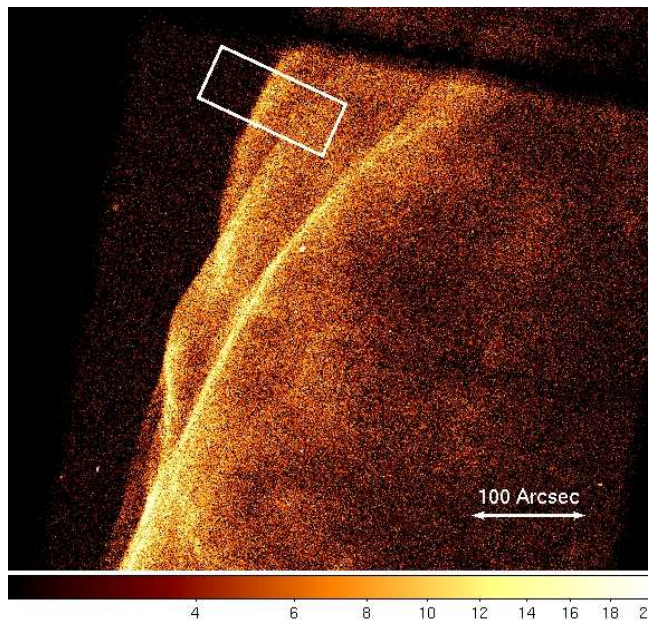}}
\caption{{\it Chandra} X-ray image of the northeastern limb showed in squared
root color scale. The white rectangle ($50''\times 120''$) is the region used
to extract the X-ray radial profile plotted in Fig.~\ref{fig:2}.}
\label{fig:1}
\end{center}
\end{figure}

\begin{figure}
\begin{center}
{\includegraphics[angle=0,width=.48\textwidth]{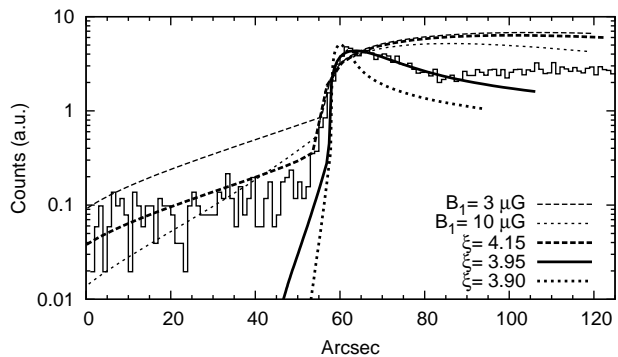}}
\caption{Radial profile of the emission in the 0.8-2.0 keV band, extracted
from the rectangular region in Fig.~\ref{fig:1} ({\it thin black line}).
Overplotted are the theoretical predictions of NLDSA for 3 different injection
efficiencies ({\it thick lines}) and those of test-particle theory for two
different values of the pre-existing turbulent magnetic field upstream ({\it
thin lines}).} 
\label{fig:2}
\end{center}
\end{figure}

Our model of NLDSA contains four free parameters, namely: the density of the
unshocked ISM, $n_0$, the shock speed, $u_0$, the proton injection efficiency,
$\xi$, and the relative normalization between electron and proton spectra,
$K_{ep}$. SN 1006 is thought to be the remnant of a type Ia SN explosion and to
expand in a fairly uniform ISM. We adopt $n_0=0.05\, \rm cm^{-3}$, based on the
measurement by \cite{Acero07} \citep[also consistent with a hadronic
interpretation of the TeV emission recently detected by H.E.S.S.;
see][]{BKV09}. We estimate the shock speed using a code for the evolution
of type Ia SNRs \citep[]{cap10}, fixing the explosion energy and the mass
of the ejecta to the standard values $E_{\rm SN}= 10^{51} \rm erg$ and $M_{eje}=
1.4 M_\odot$. Once $n_0$ is fixed, given the known age of the remnant, we
estimate $u_0=4330 {\rm km/s}$ and a radius $R_{\rm SNR}=7.7$~pc. Comparing this
latter value with the observed angular diameter of 30', the inferred distance of
SN 1006 is $d=1.77$ kpc. This makes our estimate of $u_0$ in remarkable
agreement with recent measurements of the proper motion of X-ray filaments by
\cite{Katsuda09}, which give $u_0= 4670\,(d/2\,\rm kpc)\,km/s$.

We extract the X-ray emission profile from recent {\it Chandra}
observations \citep[2008 June 24; ObsID.9107; PI: Petre; see][]{Katsuda09}.
Fig.~\ref{fig:1} shows the {\it Chandra} image of the northeastern limb. We
selected the rectangular region of $50''\times 120''$ highlighted by the white
box in the figure, so as to exclude regions where multiple shocks overlap. The
resulting radial profile, in the energy band 0.8-2.0 keV, is plotted as a
histogram in Fig.~\ref{fig:2} and is in good agreement with that of
\cite{Long03}. 

\begin{table}
\caption{\label{tab:1}}
\begin{center}
\begin{tabular}{ccccc}
\hline
$\xi$ & $B_1\, (\mu \rm G)$ & $B_2\, (\mu \rm G)$ & $R_{sub}$ &
$R_{tot}$   \\
\hline \hline
3.90     & 47  & 175  & 3.78 & 6.39 \\
3.95     & 23  & 90   & 3.93 & 5.53 \\
4.15     & 5.3 & 21   & 4.00 & 4.08 \\
$\infty$ & 3.0 & 10   & 4.00 & 4.00 \\
$\infty$ & 10 & 33   & 4.00 & 4.00 \\\hline
\end{tabular}
\end{center}
\end{table}

The curves in Fig.~\ref{fig:2} refer to different values of the injection
parameter $\xi= 3.90,\,3.95,\,4.15$ and $B_{0}=3\mu G$. The two curves labelled
as $B_{1}=3\mu G$ and $10\mu G$ are test particle cases with a turbulent
magnetic field as indicated, and with the perpendicular components compressed at
the shock, so that $B_2= \sqrt{11} B_1$. In the other cases
the magnetic field at a position $x$ in the precursor is calculated as described
by Eq.~({\ref{eq:B_amp}), and then compressed at the subshock, resulting in the
values of $B_1$ and $B_2$ reported in Table~\ref{tab:1}, together with the
compression factors $R_{sub}$ and $R_{tot}$.

It is clear that DSA in the test particle regime cannot explain either the
radial profile in the precursor or that in the downstream region, although the
latter might still be explained by requiring some magnetic damping. The
intermediate case, $\xi=4.15$, which corresponds to modest injection, is also
inappropriate for the same reasons. Moreover, the test particle and the
inefficient cases show a potentially detectable precursor, which is however not
observed. Remarkably, the solution with $\xi=3.95$ (corresponding to
$B_{2}\simeq 90\mu G$) fits well the downstream X-ray profile and shows a modest
upstream halo that is however fully consistent with the observations. For larger
injection efficiencies the upstream halo becomes smaller than the {\it Chandra}
resolution, but the predicted downstream rim is too narrow compared with
observations.

\section{Discussion and Conclusions}\label{sec:disc}

We studied the observed radial profile of the X-ray emission in SN 1006 in the
context of the theory of NLDSA. We concentrated our attention on a filament of
X-ray emission which was already studied by \cite{Long03} since it shows a
pronounced jump in emission by a factor $\sim 70$ at what appears to be the
shock surface. This filament is rather clean from the observational point of
view, in that there seems to be only one shock surface within the observed
slice. 

The conclusions of our investigation can be summarized as follows: 

1) The radial profile of the X-ray emission is very well described within the
context of NLDSA, as a consequence of a large magnetic field amplification
upstream of the shock. This scenario describes both the downstream narrow
filament and the non detection of a precursor upstream of the shock. The main
reason for this non-detection is that the magnetic field upstream becomes strong
enough to make the diffusion/loss region of electrons comparable with the
angular resolution of {\it Chandra}. Moreover, if magnetic field is amplified
upstream through the excitation of a streaming instability, the gradient in the
precursor magnetic field makes the size of the X-ray emission even smaller. 

Fixing $n_{0}=0.05\, {\rm cm}^{-3}$, we best fit the observed radial profile
with $\xi=3.95$ which implies a CR acceleration efficiency of $\sim 29$
per cent and an electron/proton ratio $K_{ep}\sim 1.3\, 10^{-4}$. An
inefficient scenario of CR acceleration appears to be inadequate to fit
observations: the narrow filaments would require some level of magnetic field
damping downstream, which however does not seem to be compatible with the
absence of filaments in the radio band \citep[]{roth}. Upstream, the inefficient
scenario typically leads to broader emission regions than observed by {\it
Chandra}.

We stress that our conclusions do not strongly depend on the assumed 
values of the upstream magnetic field and shock speed. An equally good fit of
the X-ray radial profile can be obtained for $B_0=5 \mu G$ (about the highest 
reasonable value at the galactic latitude of SN1006) and for a shock velocity of 
$4670\ {\rm km/s}$ corresponding to a larger distance of the remnant 
($d=2 {\rm kpc}$) and to an exposion energy of $1.5\cdot 10^{51}{\rm erg}$. For
a magnetic field strength $B_0=5 \mu G$ the best-fit injection parameter needs
to be changed ($\xi=3.98$), corresponding to a slightly smaller acceleration
efficiency. For a shock velocity of $4670\ {\rm km/s}$ a fit can be obtained for
$\xi=3.93$. 

2) Our conclusions are in disagreement with the point made by \cite{Long03}
that one needs a jump in the magnetic field at the shock which is $\gg 4$,
equivalent to MFA downstream. We find that the important ingredient is that
there is MFA upstream, as also found by \cite{bere2003}, although adopting
values of the ambient density and shock velocity that have now been excluded by
observations \cite[]{Acero07}. Our calculations show that in the case of SN
1006, inclusion of MFA {\it a la} \cite{belllucek} (which would further enhance
the magnetic field {\it upstream}) is not necessary.

3) Contrary to \cite{bere2003} we think that an efficient CR acceleration is not
{\it the only reasonably thinkable condition} to explain the X-ray observations.
The non detection of a CR precursor upstream can also be attributed, at least
qualitatively, to the presence of a quasi-perpendicular magnetic field, as was
also inferred by \cite{bamba03} (even if their conclusions were based on the
incorrect assumption of no magnetic jump at the shock surface); in this case CRs
can be accelerated to rather high energies \cite[]{joki87}. 
In this situation the magnetic field could be amplified downstream of the shock
as due to fluid instabilities if the shock propagates in a inhomogeneous medium 
\citep[e.g.][]{giajo07}. Whether this configuration can explain the details of
the X-ray radial profile should be subject of a dedicated investigation.

\section*{Acknowledgments}
This work was partially supported by MIUR (grant PRIN2006) and by ASI through
contract ASI-INAF I/088/06/0. This research was also supported in part by the
National Science Foundation under Grant No. PHY05-51164. We wish to acknowledge
the KITP in Santa Barbara for the exciting atmosphere during the Program
Particle Acceleration in Astrophysical Plasmas, July 26-October 3, 2009.
GM whish to thank Daniel Patnaude for helpful suggestions in handling the
{\it Chandra} data.

\end{document}